%% file: nmi_main.tex
\documentclass{article}

\usepackage[nusbwlab]{iclr2025_conference}
\usepackage{times}

\usepackage{amsmath}
\usepackage{amssymb}
\usepackage{amsfonts}
\input{math_commands.tex}   

\usepackage{hyperref}       
\usepackage{url}            
\usepackage{booktabs}       
\usepackage{nicefrac}       
\usepackage{microtype}      
\usepackage{xcolor}         
\usepackage{fancyhdr} 
\usepackage{colortbl}
\usepackage{enumitem}
\usepackage{algorithm,algorithmic}
\usepackage{tcolorbox}
\usepackage{graphicx}       
\usepackage{placeins}       
\definecolor{backred}{RGB}{255, 190, 190}
\definecolor{backblue}{RGB}{220, 230, 250}

\definecolor{tbdpink}{RGB}{214, 51, 132}
\definecolor{provblue}{RGB}{0, 90, 200}
\definecolor{bwlblue}{RGB}{0,114,206} 
\graphicspath{{nmi/figures/}{figures/}}

\providecommand{\bwlheadtext}{Preprint}

\title{Chemical Chain-of-Thought Functions as a Hallucination-Prone Molecular Scratchpad 
}

\author{Jiatong Li$^{1,2}$ \quad Yuxuan Ren$^{1}$ 
\quad Weida Wang$^{3}$ 
\quad Xiao-yong Wei$^{2}$ 
\quad Yatao Bian$^{1\dagger}$ \\
$^1$National University of Singapore \\
$^2$Hong Kong Polytechnic University \\
$^3$Shanghai Artificial Intelligence Laboratory\\
$^\dagger$Corresponding to: \texttt{ybian@nus.edu.sg}}

\begin{document}
\maketitle

\input{nmi/sections/abs}

\input{nmi/sections/introduction}
\input{nmi/sections/results}
\input{nmi/sections/discussion}

\input{nmi/sections/methods}


\input{nmi/sections/availability}


\bibliography{references}
\bibliographystyle{iclr2025_conference}

\end{document}

%% file: math_commands.tex
\usepackage{amsmath,amsfonts,bm}

\def\eqref#1{equation~\ref{#1}}

\def\1{\bm{1}}

\DeclareMathAlphabet{\mathsfit}{\encodingdefault}{\sfdefault}{m}{sl}
\SetMathAlphabet{\mathsfit}{bold}{\encodingdefault}{\sfdefault}{bx}{n}



%% file: nmi/sections/abs.tex
\begin{abstract}
Chemical reasoning language models are expected to derive molecular answers through faithful chain-of-thought (CoT).
However, across four reasoning model families and twelve chemistry tasks, hallucination is widespread and largely decoupled from answer correctness: correct answers often coexist with fabricated structural claims absent from the relevant molecules. 
Yet this does not make the reasoning trace computationally irrelevant. 
Attribution analyses suggest a shared scratchpad function expressed in model-specific forms: Chem-R and ether-0 rely on fragmented SMILES drafts, whereas ChemDFM-R emphasizes scaffold, positional, and naming cues. 
Notably, perturbing Chem-R's SMILES sketches degrades generation, showing that structural drafts can be causally load-bearing even when verbal structural claims are largely inert. 
Together, these results show that chemical CoT is neither a faithful explanation nor merely a post-hoc rationalization, but a hallucination-prone molecular scratchpad. 
This finding cautions against treating CoT as direct evidence of faithful reasoning and motivates process-level supervision beyond answer-only evaluation.
\end{abstract}

%% file: nmi/sections/introduction.tex
\section{Introduction}
\label{sec:intro}

Chain-of-thought (CoT) reasoning~\citep{wei2022chainofthought,yeo2025demystifying,chen2026towards} has become a widely adopted paradigm for eliciting intermediate reasoning process in large language models (LLMs).
In science, it is appealing because a model returns not only an answer but also a rationale that a human expert can inspect~\citep{zhao2025molreasoner,li2025molr1,wang2025chemr,zhao2025chemdfm}. 
However, whether that rationale can be trusted is difficult to test. 
In natural-language reasoning, claims may be fact-checkable, but intermediate rationales are rarely grounded in a formal object that permits systematic verification~\citep{jacovi2020towards,lyu2023faithful,turpin2023language,lanham2023measuring}.
Chemistry provides such a referent: when a model claims that a molecule contains a carboxylic acid or retains an amide, the statement can be checked directly against the molecular graph~\citep{li2024tomg}. 
Chemical reasoning therefore offers CoT traces whose explicit claims can be verified against molecular structures, providing a direct test of whether these models reason as their explanations suggest.

Here we treat chemical rationales as collections of testable molecular structural claims. 
We mainly focus on functional-group-level claims, which are frequent in chemical rationales and can be evaluated reliably by exact structure matching.
We developed a molecular claim-grounding framework that parses each response into a reasoning trace and a final answer, extracts chemical claims from the trace, and checks each claim against the task-relevant structures. 
The framework isolates a particularly diagnostic failure: a functional group asserted in the trace but absent from the input, predicted molecule and reference.
We term this an extrinsic reasoning fabrication (ER), which directly exposes a mismatch between the model's stated reasoning and the task-relevant molecular structures.

Applied to four reasoning model families~\citep{guo2025deepseek,zhao2025chemdfm,wang2025chemr,narayanan2025ether0} across twelve generative molecular tasks, the framework shows that these hallucinations are widespread, arising even in responses whose final molecule is exactly correct. 
Taking text-to-molecule generation on the ChEBI-20 benchmark~\citep{edwards2022translation,edwards2021text2mol} as an example, $35\%$ of Chem-R~\citep{wang2025chemr} reasoning traces contain at least one fabricated functional-group claim. Specifically, $13\%$ of Chem-R responses on this task are both exactly correct and accompanied by at least one such fabrication. Moreover, fabrication is nearly uncorrelated with correctness, with the absolute Pearson correlation between the fabrication score and exact match below $0.02$.
Thus, a model can get the molecule right while misstating the chemistry used to justify it. 
Answer-level metrics cannot reveal this mismatch: they score the molecular output but leave the stated chemical reasoning path unobserved. 
The verifiable content of the reasoning traces can therefore be decoupled from the correctness of the final answer.

Direct intervention explains how the two coexist. Corrupting a verified functional-group claim inside the reasoning trace flips only $1$--$11\%$ of originally-correct answers, similar to a synonym control, whereas applying the same corruption to the input has a much larger effect, flipping $16$--$78\%$.
This asymmetry suggests that the model relies more on input-grounded evidence than on its own functional-group claims.
However, the trace is not disposable. Removing it or substituting another problem's trace changes generation and significantly influences the performance. 
Therefore, the trace itself may still carry answer-relevant information even though its explicit claims are unreliable. 

One possible interpretation is that the useful content of chemical chain-of-thought lies less in its verbal explanations than in its function as a molecular scratchpad. 
Before producing the final answer, the model writes intermediate structural cues, such as fragments, positions, or ring systems, that remain available during answer generation. 
The complete target structure is absent from most trace, and incomplete structural drafts can nevertheless support generation.
The form of this scratchpad is model-specific. Chem-R~\citep{wang2025chemr} and ether-0~\citep{narayanan2025ether0} draft fragmented SMILES structures that affect their answers, whereas attribution analyses suggest that ChemDFM-R~\citep{zhao2025chemdfm} places more weight on scaffold-level, positional and nomenclature cues. Perturbing the SMILES sketches degrades Chem-R's generation, while perturbing functional-group claims has little effect.

Taken together, these results suggest that chemical CoT is neither a faithful explanation~\cite{zhao2025molreasoner,zhao2025chemdfm} nor merely a post-hoc rationalization~\citep{cox2026decoding,arcuschin2025chain}, but a hallucination-prone molecular scratchpad. Its readable prose is often unsupported, whereas structural drafts can contribute to generation. Answer-only evaluation cannot distinguish traces that reach the same answer through differently grounded claims. To test process-level supervision, we continued training from the released Chem-R checkpoint using a verification-grounded reward that pays the answer-accuracy term only for traces without ER fabrications. We refer to the resulting model as Chem-R-Faithful, or simply Faithful throughout this paper. This intervention reduces ER fabrications while preserving reported performance.

%% file: nmi/sections/results.tex
\section{Results}
\label{sec:results}

\begin{figure}[t]
\centering
\includegraphics[width=\linewidth]{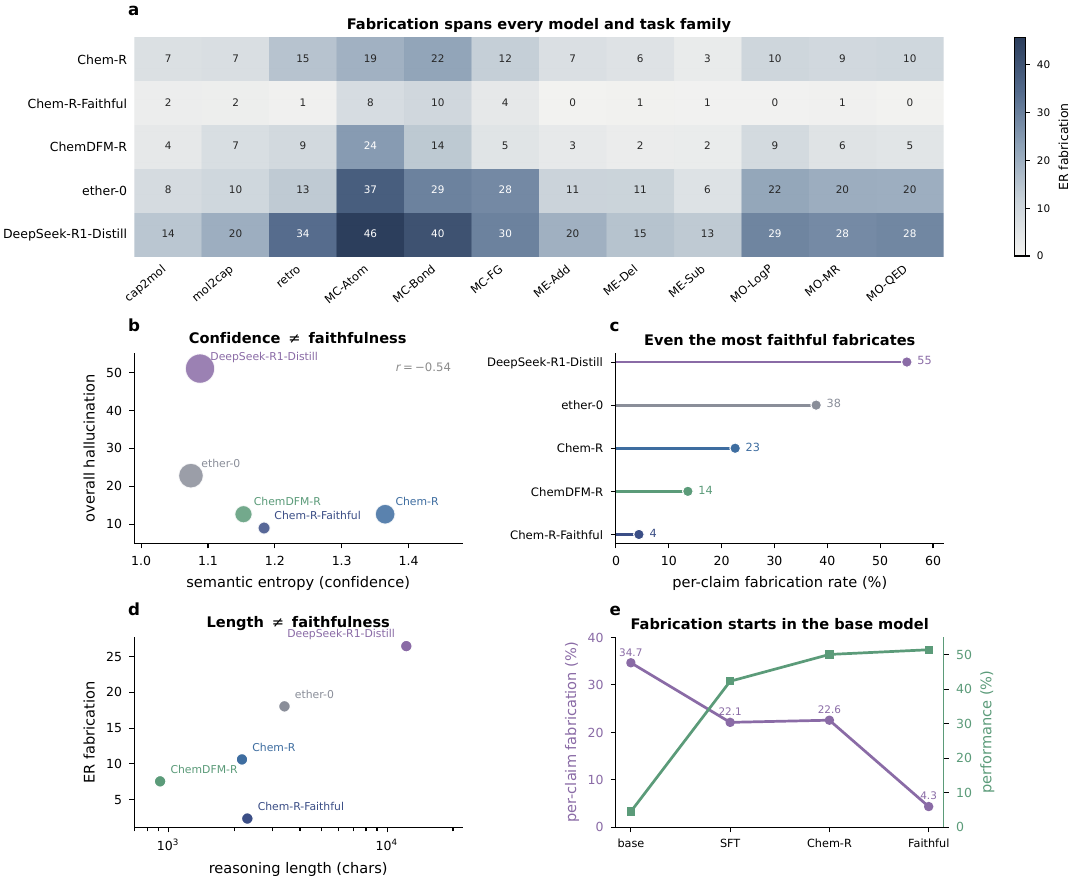}
\caption{\textbf{Reasoning hallucination is pervasive and not decided by confidence or length.} 
\textbf{a}, Per-response ER, a bounded fabrication score (0--100) that increases with the fraction of a trace's claimed functional groups that are unsupported, averaged over traces, across the four released models and Chem-R-Faithful on twelve generative task variants.
\textbf{b}, Each released model and Chem-R-Faithful as one point in (semantic entropy, overall hallucination), with marker area increasing with ER; all three quantities are averaged over the twelve generative task variants. Confidence does not separate faithfulness from fabricating models.
\textbf{c}, Among the released models, even the most faithful fabricates approximately $14\%$ of its claims; our verification-grounded Chem-R-Faithful is the exception, reducing the rate to $4.3\%$. \textbf{d}, ER likewise does not follow reasoning length: the longest reasoner is the least faithful, whereas Chem-R-Faithful has the lowest ER.
\textbf{e}, Training lineage from the pre-chemistry-SFT base model through chemistry supervised fine-tuning with chain-of-thought reasoning (SFT) and answer-only GRPO (Chem-R) to verification-grounded GRPO. The final transition continues training from the released Chem-R checkpoint and yields Chem-R-Faithful. Per-claim fabrication (purple) is highest before chemistry SFT and falls to $4.3\%$ only after the verification-grounded stage, while task performance (green) is preserved.
}
\label{fig:pervasive}
\end{figure}

We first formalize how explicit molecular claims can be checked
(Fig.~\ref{fig:framework}), then measure their fabrication and relation to answer
correctness (Figs.~\ref{fig:pervasive} and~\ref{fig:decoupling}). Subsequent
experiments test whether verbal claims or drafted structures affect the answer and
characterize model-specific scratchpad forms
(Figs.~\ref{fig:mechanism}--\ref{fig:scratchpad}).

\subsection{Reasoning hallucination is a chemically checkable grounding failure}
\label{sec:res-framework}

\begin{figure}[t]
\centering
\vskip -0.2in
\IfFileExists{nmi/figures/fig1_framework_white.pdf}{%
  \includegraphics[width=\linewidth]{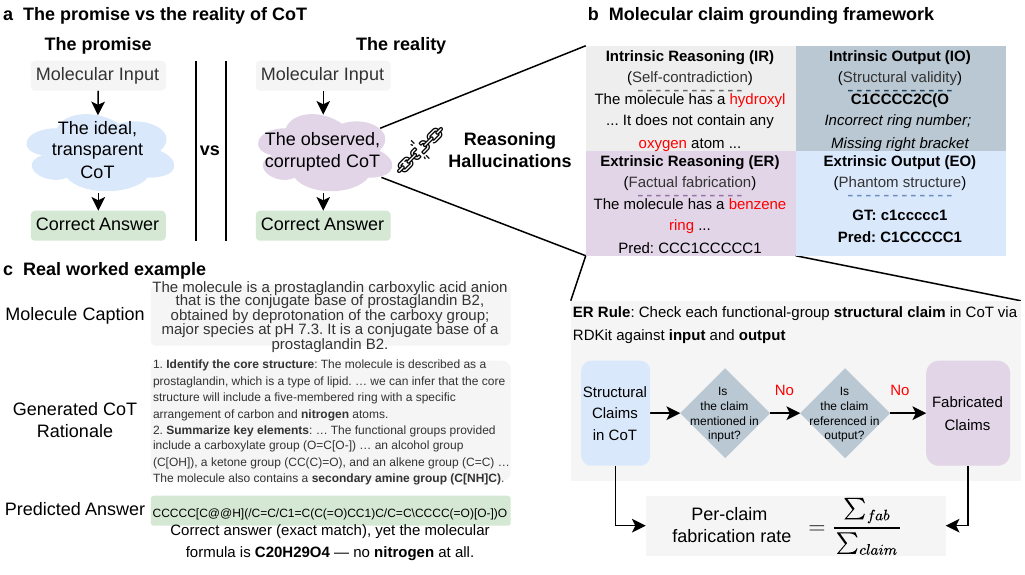}%
}{%
  \fbox{\parbox[c][0.25\textheight][c]{0.94\linewidth}{\centering
  Missing source figure: \texttt{nmi/figures/fig1\_framework\_white.pdf}}}%
}
\par\vspace{0.35em}
\IfFileExists{nmi/figures/fig1_measures.pdf}{%
  \includegraphics[width=\linewidth]{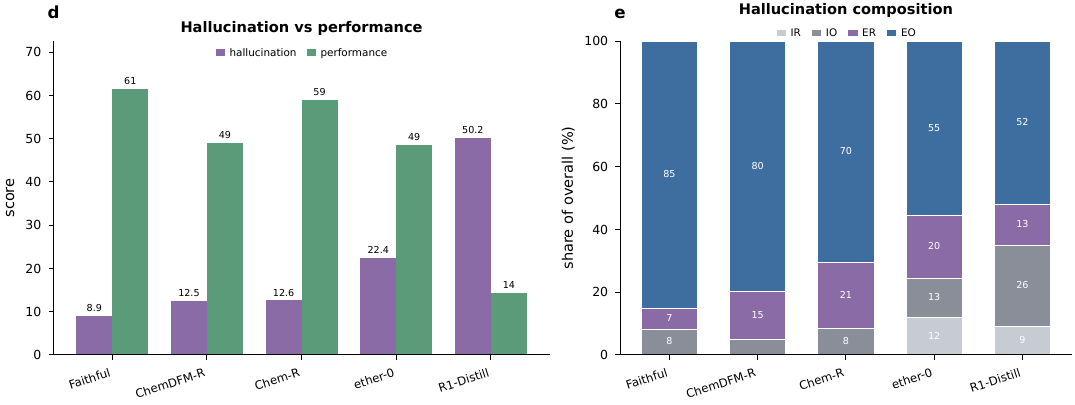}%
}{%
  \fbox{\parbox[c][0.14\textheight][c]{0.94\linewidth}{\centering
  Missing source figure: \texttt{nmi/figures/fig1\_measures.pdf}}}%
}
\vskip -0.2in
\caption{
\textbf{Chemical chain-of-thought can contain structurally verifiable hallucinations even when the answer is correct.} \textbf{a}, The promise of transparent chain-of-thought (CoT) is contrasted with the observed failure mode: a corrupted rationale can still lead to the correct answer. \textbf{b}, Molecular claim-grounding framework. Crossing the location of an error (reasoning or output) with its grounding (intrinsic or extrinsic) yields intrinsic reasoning (IR), intrinsic output (IO), extrinsic reasoning (ER) and extrinsic output (EO) errors. The schematic illustrates these categories with a self-contradictory reasoning claim, an invalid output SMILES, a fabricated benzene-ring claim and a phantom output structure, respectively. For ER, each functional-group claim in the CoT is checked with RDKit against the input, the predicted output and, where the task defines one, the reference; a claim supported by none of them is labelled fabricated, and the per-claim fabrication rate is the fraction of structural claims so labelled.
\textbf{c}, Worked caption-to-molecule (cap2mol) example. The CoT invents nitrogen atom, whereas the predicted SMILES is exactly correct and contains no nitrogen, exposing a reasoning fabrication invisible to answer-only evaluation.
\textbf{d}, Aggregate hallucination score and task performance (both on a 0--100 scale; unweighted means over the twelve task variants) for the four released reasoning models and Chem-R-Faithful (``Faithful'').
\textbf{e}, Within-model percentage composition of aggregate hallucination across IR, IO, ER and EO; each stacked bar is normalized to 100\%.}
\label{fig:framework}
\end{figure}

Chemical reasoning traces often contain structural claims that can be verified against the molecular structure.
Here we anchor the analysis on functional-group claims, which are frequent in chemical rationales and reliably decidable by RDKit SMARTS matching~\citep{landrum2013rdkit}. 
Our molecular claim-grounding framework parses a response into a reasoning trace and a final answer, extracts the chemical entities and transformations asserted in the trace, and checks each against the task-relevant structures (Fig.~\ref{fig:framework}). 
The framework separates two axes that answer-level metrics conflate: whether an error appears in the reasoning or in the final answer (i.e., output), and whether a claim contradicts an external reference or the model's own generated content. 
The most diagnostic cell is \emph{extrinsic reasoning fabrication} (ER): a functional group asserted in the trace that is absent from the input and the predicted molecule alike, and from the reference structure where the task defines one.
Because such a group is present in neither the question nor any answer, the claim is not merely a disagreement with an imperfect reference. 
We report the per-response ER score and, as finer probes, claim precision (the fraction of structural claims that verify) and the count of grounded claims.

Figure~\ref{fig:framework}c illustrates why a separate reasoning metric is needed.
Chem-R returns the exact-matched molecule but invents nitrogen atoms in its trace, although the predicted molecule contains no nitrogen at all.
Only a reasoning-side check exposes the fabrication.

At the model level, hallucination does not simply mirror task performance (Fig.~\ref{fig:framework}d). Chem-R and ChemDFM-R have nearly identical aggregate hallucination scores despite a ten-point performance gap, whereas Chem-R-Faithful combines the lowest hallucination with the highest performance. The four error classes also contribute differently across models (Fig.~\ref{fig:framework}e): output-side errors account for most of the aggregate score, but ER remains present in every model, motivating the reasoning-side analyses below.

\subsection{Reasoning hallucination is pervasive and not decided by confidence or length}
\label{sec:res-pervasive}

Applying the framework to four open-source reasoning models and the Chem-R-Faithful checkpoint across twelve generative task variants shows that fabrication is common (Fig.~\ref{fig:pervasive}a).
Chemistry-specialized models score substantially lower on both overall hallucination and ER than the general reasoners, but none are clean. Among the released models, mean ER across tasks ranges from $7.6$ to $26.4$. Grounded process reward lowers Chem-R-Faithful to $2.4$ but does not eliminate fabrication. 

Two intuitive proxies fail to explain the pattern. Semantic entropy, a sampling-based uncertainty measure used to detect hallucination~\citep{kuhn2023semantic,farquhar2024detecting}, spans only approximately $1.07$--$1.37$ while hallucination varies several-fold. Across the five models its descriptive correlation also opposes a simple confidence account (Fig.~\ref{fig:pervasive}b). 
Meanwhile, reasoning length is not necessarily related to faithfulness: although models generally exhibit a ``more reasoning, more fabrication'' pattern, Chem-R-Faithful substantially reduces hallucinations while preserving reasoning length (Fig.~\ref{fig:pervasive}d). 

By contrast, the per-claim fabrication rate tracks the released-model ranking ( Fig.~\ref{fig:pervasive}c), showing that even the most faithful released model fabricates about one claim in seven. 
Our Chem-R-Faithful checkpoint is the exception, reducing the rate to $4.3\%$.

A comparison along one Chem-R lineage further suggests that hallucination predates chemistry-specific fine-tuning: the per-claim fabrication rate is highest in the original backbone model ($34.7\%$), is reduced by supervised fine-tuning ($22.1\%$), is not further reduced by answer-only reinforcement learning ($22.6\%$), and falls to $4.3\%$ only once the reward grounds the trace's claims, even as task performance saturates earlier (Fig.~\ref{fig:pervasive}e).

\subsection{Answer correctness is a poor proxy for extrinsic reasoning hallucination}
\label{sec:res-decoupling}

\begin{figure}[t]
\centering
\includegraphics[width=\linewidth]{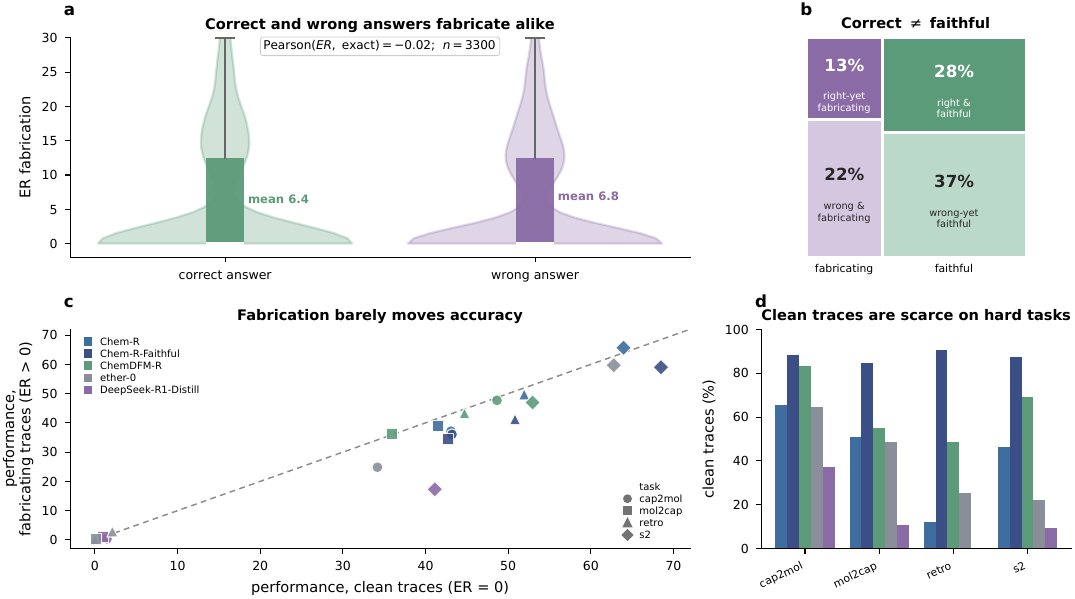}
\caption{\textbf{Fabrication is decoupled from answer correctness.} \textbf{a}, For Chem-R on cap2mol ($n=3300$), the ER distributions of correct and wrong answers nearly coincide (means $6.4$ versus $6.8$; $|\mathrm{Pearson}|<0.02$). 
\textbf{b}, The same Chem-R cap2mol responses as a joint distribution of correctness and faithfulness: right-yet-fabricating and wrong-yet-faithful cases are both common. 
\textbf{c}, Across the four released models and Chem-R-Faithful, performance on clean ($\mathrm{ER}=0$) and fabricating ($\mathrm{ER}>0$) traces lies close to the equality line across task families. 
\textbf{d}, The clean-trace rate (ER $=0$) varies widely: it is lowest for the general reasoners and on difficult tasks for the released chemistry models, but reaches approximately $84$--$91\%$ for Chem-R-Faithful.}
\label{fig:decoupling}
\end{figure}

Answer correctness provides little information about whether a reasoning trace contains extrinsic hallucinations. 
On the full caption-to-molecule (cap2mol) test set for Chem-R ($n=3300$), the ER distributions of exactly correct and incorrect responses are nearly indistinguishable (Fig.~\ref{fig:decoupling}a), with mean ER scores of $6.4$ and $6.8$, respectively. 
Consistently, the correlation between ER and the exact-match indicator is close to zero ($|\mathrm{Pearson}(\mathrm{ER},\text{exact match})|<0.02$; Fig.~\ref{fig:decoupling}a). Similarly weak associations are observed for mol2cap ($-0.04$) and retrosynthesis ($-0.05$).

The joint distribution makes the distinction between answer correctness and trace groundedness concrete. 
Of all Chem-R's cap2mol responses, $28\%$ are both exactly correct and clean ($\mathrm{ER}=0$), whereas $13\%$ are exactly correct but contain at least one detected fabrication ($\mathrm{ER}>0$).
Thus, although the overall exact-match rate is $41\%$, approximately $31\%$ of exactly correct responses contain a fabrication. 
Conversely, the majority of incorrect responses still have clean traces (Fig.~\ref{fig:decoupling}b). 
Correctness therefore neither guarantees a clean trace nor does incorrectness imply a fabricated one.

The same qualitative pattern appears at the task-family level. 
When official task performance is computed separately for clean and fabricating traces, the resulting points generally remain near the equality line (Fig.~\ref{fig:decoupling}c). 
On the S$^2$-Bench~\citep{li2024tomg} tasks, Chem-R is even marginally more accurate among responses whose traces contain fabrications, further illustrating that the relation is not monotonic. 
This decoupling should not be interpreted as implying that clean traces are universally common: across these models, especially on harder tasks in retrosynthesis and S$^2$-Bench, their absolute prevalence is often low. 
By contrast, Chem-R-Faithful raises the clean-trace rate to approximately $84$--$91\%$ across the four task families (Fig.~\ref{fig:decoupling}d). 

Answer-level metrics are therefore insufficient for evaluating the reasoning trace. 
They assess the molecular output, but not whether the chemical claims used to justify that output are supported. 
The near-zero associations in Fig.~\ref{fig:decoupling} show that answer correctness is a poor proxy for claim-level groundedness: a correct answer may contain fabricated reasoning, while an incorrect answer may still have a trace with no detected fabrications.

\subsection{Functional-group-level claims in the reasoning trace exert little causal influence on the final answer}
\label{sec:res-mechanism}

How can a correct answer coexist with a reasoning trace that contains a fabricated claim? One possibility is that the verbal claims in the trace do not materially determine the final answer. We tested this possibility using targeted interventions on functional-group mentions.

\begin{figure}[t]
\centering
\includegraphics[width=\linewidth]{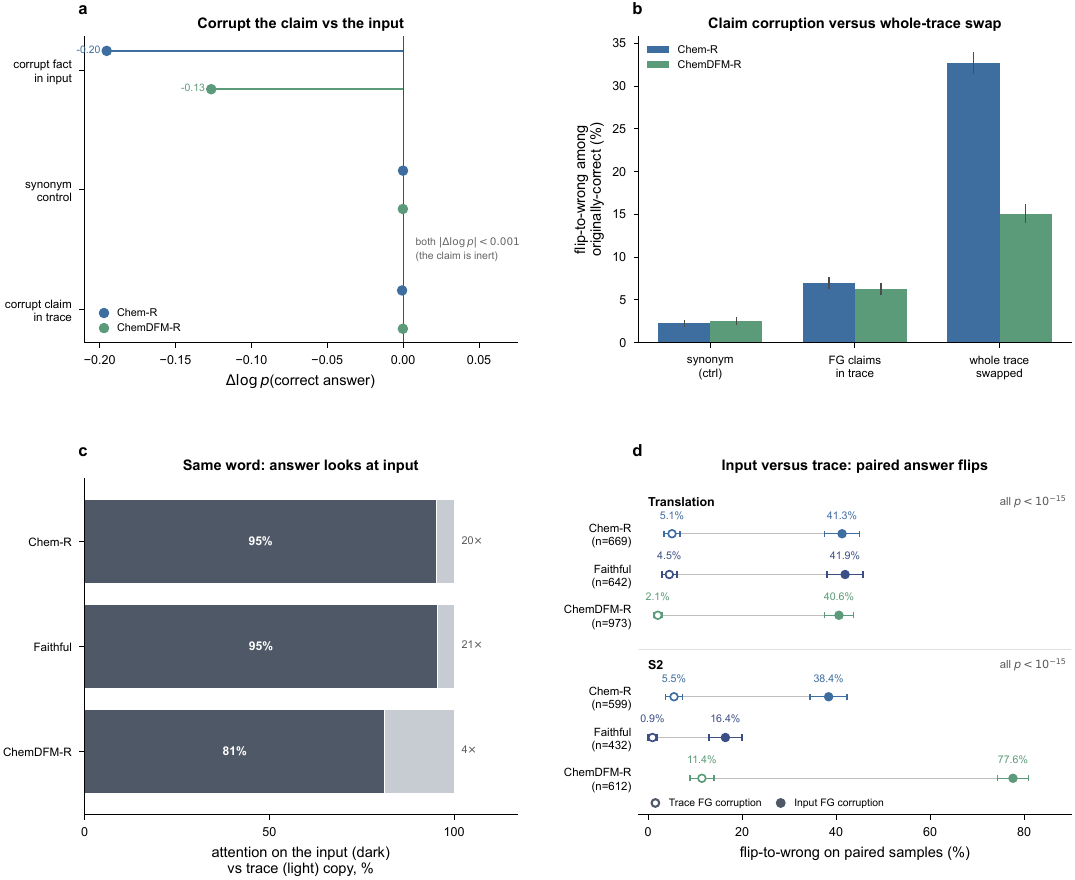}
\caption{\textbf{The answer is more sensitive to functional-group information in the input than in the reasoning trace.}
\textbf{a}, Mean teacher-forced change in correct-answer log-probability on caption-to-molecule. Sample sizes for trace-claim corruption, synonym substitution and input corruption are $n=924/919/840$ for Chem-R and $1297/1281/1199$ for ChemDFM-R.
\textbf{b}, Flip-to-wrong rates among originally-correct responses in the molecule--caption translation and retrosynthesis tasks. For synonym substitution, trace-claim corruption and whole-trace replacement, respectively, $n=4936/4950/4950$ for Chem-R and $4149/4163/4163$ for ChemDFM-R.
\textbf{c}, Relative shares of mean middle-layer attention to matched input and trace occurrences of the same functional-group term on caption-to-molecule (n = 265 matched input–trace comparisons for Chem-R, 233 for Chem-R-Faithful and 255 for ChemDFM-R).
\textbf{d}, Paired trace (open circles) and input (filled circles) corruption, separately for translation and S2. Connected points use the same originally-correct responses with both results available ($n$ shown). S2 pools functional-group customization and component addition, deletion and substitution. Across these three models and two task groups, input flip rates span $16.4$--$77.6\%$; the upper endpoint is ChemDFM-R on S2 ($475/612$). Paired differences span $15.5$--$66.2$ percentage points (all six two-sided exact McNemar tests, $p<10^{-15}$, unadjusted). Error bars in \textbf{b},\textbf{d} are $95\%$ normal-approximation confidence intervals.}
\label{fig:mechanism}
\end{figure}

Starting from clean caption-to-molecule traces, we selected a functional-group mention that was verified to be present in the target molecule and replaced it with a random incorrect group. 
We then measured the teacher-forced log-probability of the original correct answer. As a matched comparison, we applied the same corruption to the functional-group in the input caption;
synonym substitutions also served as a meaning-preserving control. 
Corrupting the trace had a negligible effect on the answer log-probability ($\Delta\log p=-0.0007$ for Chem-R), comparable to the synonym control. 
By contrast, corrupting the input produced reductions in log-probability that were approximately two orders of magnitude larger in absolute value ($-0.20$ for Chem-R and $-0.13$ for ChemDFM-R; Fig.~\ref{fig:mechanism}a).



A complementary answer-flip test showed the same asymmetry. Among the Chem-R responses that were originally correct on the three translation tasks ($n = 4{,}950$), corrupting the functional-group claims in the trace flipped only 7\% of answers to incorrect (5.6\% on cap2mol alone), close to the synonym control (2.3\%; Fig.~\ref{fig:mechanism}b). For the input-versus-trace comparison, we restricted the analysis to the 669 originally correct responses for which results were available under both interventions. On this paired subset, trace corruption flipped 5.1\% (34/669) of answers, compared with 41.3\% (276/669) for input corruption (Fig.~\ref{fig:mechanism}d); the 7\% and 5.1\% trace rates therefore summarize different response sets rather than different calculations on the same set. Swapping the entire trace, by contrast, changed about 30\% of previously correct answers, 
indicating that, although functional-group claims are not causally load-bearing, the trace as a whole remains functionally coupled to the final answer (Fig.~\ref{fig:mechanism}b). Across models and tasks, the answer-flip rate was 1--11\% for trace corruption and 16--78\% for input corruption, yielding paired gaps of 15--66 percentage points ($p < 10^{-15}$).

The weak effect of trace corruption persisted both for claims restated from the input ($\Delta\log p=-0.0092$ to $-0.0000$) and for claims derived by the model ($-0.0013$ to $-0.0004$). 
Thus, the result cannot be explained solely by trace claims redundantly repeating information from the input. 
Attention patterns provide complementary descriptive evidence: when a functional-group term appeared in both the input and the trace, answer tokens assigned $20\times$ more attention to its input occurrence for Chem-R and $4\times$ more for ChemDFM-R (Fig.~\ref{fig:mechanism}c). 

Together, these interventions indicate that, for the functional-group claims tested here, the final answer is substantially more sensitive to information in the input than to the corresponding verbal claim in the reasoning trace.
This asymmetry seems consistent with parts of the trace serving as post-hoc rationalization~\citep{turpin2023language,lanham2023measuring}, but it does not establish that every claim, or the trace as a whole, is generated post hoc.

\subsection{SMILES drafts can be causally load-bearing in the Chem-R lineage}
\label{sec:res-draft}

\begin{figure}[t]
\centering
\includegraphics[width=\linewidth]{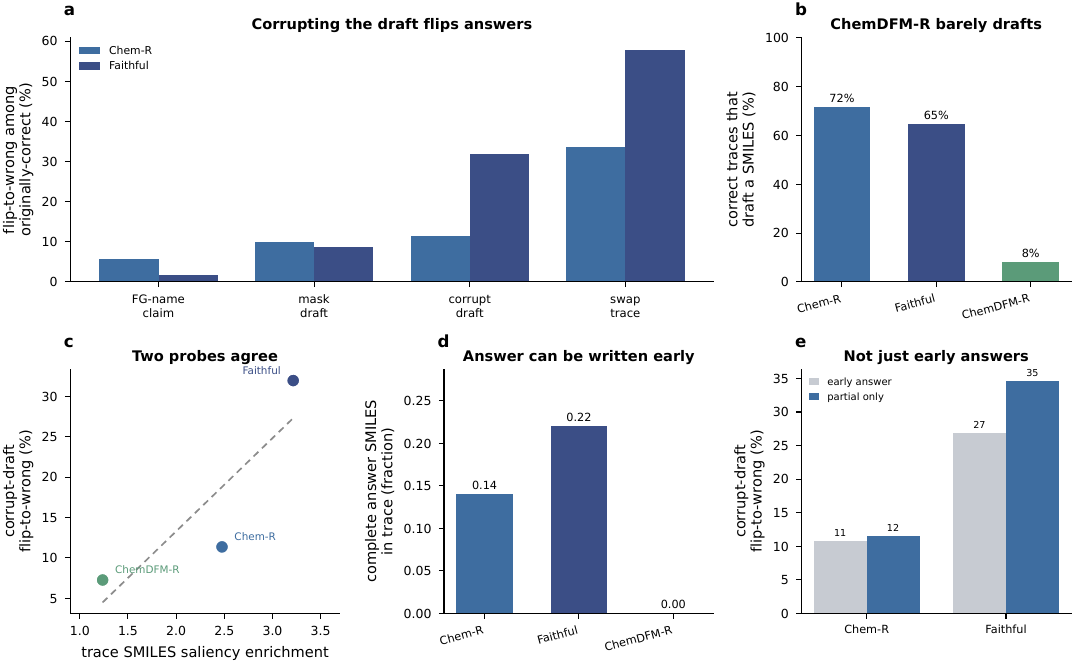}
\caption{\textbf{SMILES drafts affect generation in the Chem-R lineage.} \textbf{a}, Flip-to-wrong rate among originally-correct answers that contain a drafted SMILES (all conditions scored on this same subset) under trace perturbations. Corrupting the drafted SMILES flips more answers than corrupting functional-group names or masking the draft for both Chem-R and Chem-R-Faithful.
\textbf{b}, ChemDFM-R drafts a SMILES in only $8\%$ of correct traces and serves as a negative control for the role of drafted SMILES. 
\textbf{c}, Across models, the corrupt-draft flip rate follows the SMILES-saliency enrichment of Fig.~\ref{fig:scratchpad}b.
\textbf{d}, The complete answer SMILES appears within the trace (early or complete answering) in $14$--$22\%$ of Chem-R and Chem-R-Faithful traces, and in approximately zero ChemDFM-R traces.
\textbf{e}, The corrupt-draft effect is not an early-answer artefact: among draft-writing traces that contain only partial structure and never the complete answer ($77\%$/$65\%$ for Chem-R/Chem-R-Faithful), corrupting the draft flips $11.5\%$/$34.6\%$ of answers, close to the rates on the full draft-writing subset.}
\label{fig:draft}
\end{figure}

We argue that the trace is not entirely post-hoc. Although corrupting its functional-group names has little effect (Fig.~\ref{fig:mechanism}), replacing the whole trace with another molecule's trace raises answer entropy by $0.05$--$0.32$; corrupting only the functional-group names changes it by $0.03$--$0.04$ for the three Llama-family models. 
Some answer-relevant information therefore remains outside the verbal claims. A prominent non-verbal element in Chem-R traces is the drafted SMILES. We tested this structural component directly by perturbing only the SMILES written in the trace, leaving the functional-group prose intact and regenerating the answer on originally-correct responses that contain such a draft (Fig.~\ref{fig:draft}). 
Every condition was scored on this same draft-writing subset. Replacing the draft with a valid-but-wrong structure flips more answers than corrupting the functional-group claims: $11.4\%$ versus $5.7\%$ for Chem-R ($2\times$) and $31.9\%$ versus $1.8\%$ for Chem-R-Faithful ($17\times$; Fig.~\ref{fig:draft}a). 
Corruption also flips more answers than masking the draft, indicating that the replacement redirects generation rather than merely removing useful context.

The result is not explained by complete answers being written early in the trace. Although we do find that the complete SMILES strings appear in $14$--$22\%$ of Chem-R and Chem-R-Faithful traces (Fig.~\ref{fig:draft}d), most draft-writing traces contain only partial structure ($77\%$ for Chem-R and $65\%$ for Chem-R-Faithful). On this partial-only subset, corrupting the draft still flips $11.5\%$ and $34.6\%$ of answers, respectively, close to the rates on the full draft-writing subset (Fig.~\ref{fig:draft}e).

ChemDFM-R provides a negative control for the role of drafted SMILES. It drafts a SMILES in only $8\%$ of correct traces, and corrupting that draft flips $7.3\%$ of originally correct answers (Fig.~\ref{fig:draft}b). Across the three models, the corrupt-draft flip rate follows the independently measured SMILES-saliency enrichment reported in the next section (Fig.~\ref{fig:draft}c). This three-point comparison is descriptive, but its agreement with the perturbation results supports the localization of a load-bearing SMILES component in the Chem-R lineage. The tested draft effects remain smaller than the input effect, for which corruption flips $16$--$78\%$ of answers (Fig.~\ref{fig:mechanism}). 
Thus, the trace can serve as a working draft while the input remains the primary source of the answer.

\subsection{Attribution analyses suggest model-specific forms of the molecular scratchpad}
\label{sec:res-scratchpad}

\begin{figure}[t]
\centering
\includegraphics[width=\linewidth]{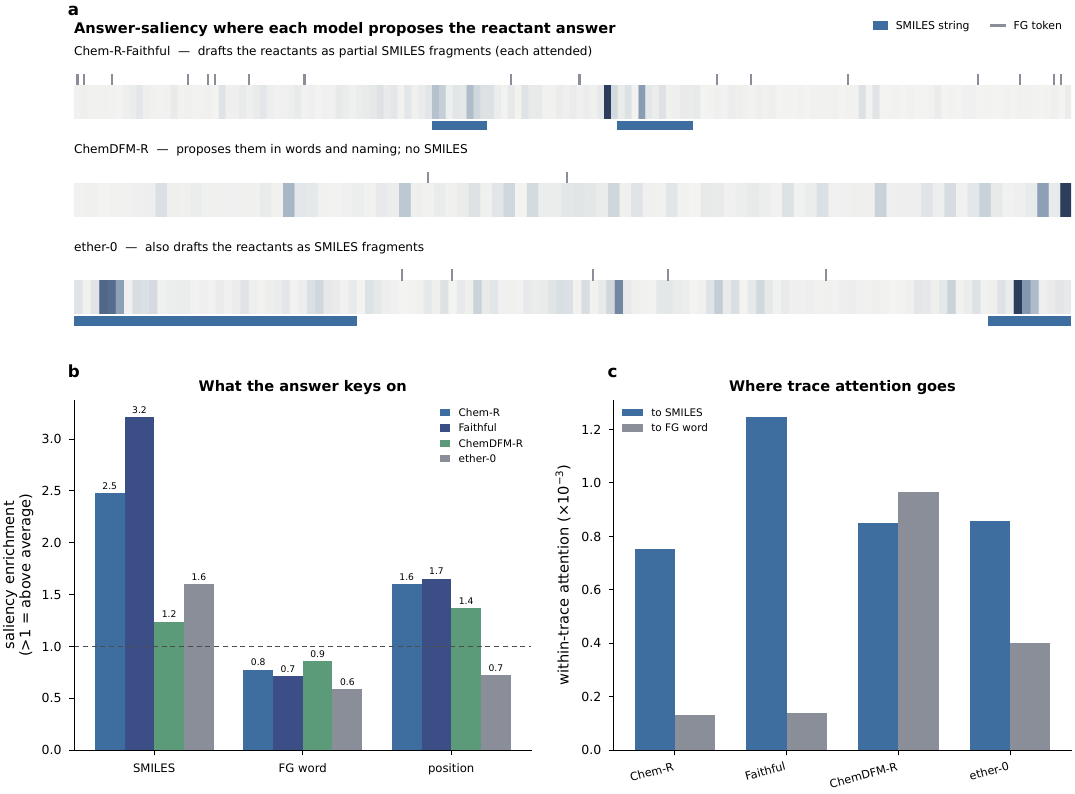}
\caption{\textbf{Attribution analyses suggest model-specific forms of the molecular scratchpad.} 
\textbf{a}, Per-token answer-saliency in the region of a retrosynthesis trace where each model proposes the reactant answer (colour, answer-saliency; shaded spans mark drafted SMILES fragments, ticks functional-group tokens). 
Chem-R-Faithful drafts the reactants as two dispersed partial SMILES fragments with high answer-saliency; ether-0 also drafts SMILES fragments, whereas ChemDFM-R writes no SMILES and instead expresses the reactants through positional and naming cues.
\textbf{b}, Per-token answer-saliency enrichment by trace token type (values $>1$ carry more answer-sensitivity than the average trace token): SMILES fragments are most enriched for Chem-R and Chem-R-Faithful ($2.5$--$3.2\times$), remain enriched for ether-0 ($1.6\times$), and are weakest for ChemDFM-R ($1.2\times$). Functional-group words are depleted ($<1$) for all four checkpoints.
\textbf{c}, Within-trace attention from the answer favours SMILES fragments for Chem-R, Chem-R-Faithful and ether-0, but reverses to functional-group words for ChemDFM-R.}
\label{fig:scratchpad}
\end{figure}

Having established that drafted SMILES can affect generation in the Chem-R lineage, we next asked whether the same scratchpad form is prominent across models. We use gradient$\times$input saliency~\citep{ancona2018towards} to measure the answer's sensitivity to each trace token in Chem-R, Chem-R-Faithful, ChemDFM-R and ether-0 (Fig.~\ref{fig:scratchpad}).

For Chem-R and Chem-R-Faithful, the most salient token type per token is the SMILES fragments (enrichment $2.5\times$ and $3.2\times$ above the average trace token), while functional-group words are depleted ($0.71$--$0.77\times$;
Fig.~\ref{fig:scratchpad}b). Within-trace attention shows the same preference: answer tokens attend more to SMILES fragments than to functional-group words (Fig.~\ref{fig:scratchpad}c). These convergent attribution measures independently localize the SMILES component tested in Fig.~\ref{fig:draft}, similar to the use of generated tokens as an intermediate-computation scratchpad~\citep{nye2021show}.

Meanwhile, ether-0 shows a weaker but still SMILES-oriented profile. Its example trace drafts the reactants as SMILES fragments; across traces, SMILES tokens are enriched $1.6\times$ in answer saliency while functional-group words are depleted to $0.6\times$, and within-trace attention to SMILES is about twice that to functional-group words (Fig.~\ref{fig:scratchpad}a--c). This suggests that the attribution asymmetry is not limited to the Chem-R lineage.

In this limited sense, ChemDFM-R exhibits a more conservative scratchpad profile, with less reliance on explicit SMILES fragments and a more distributed use of chemically descriptive cues.
SMILES fragments account for little trace saliency ($1.0\%$ of the total mass; $1.2\times$ per-token enrichment), and within-trace attention favours functional-group terms over SMILES (Fig.~\ref{fig:scratchpad}c). 
Its answer-sensitive spans instead contain scaffold, positional, and nomenclature cues, suggesting that answer-relevant information is distributed across chemically descriptive language rather than concentrated in an explicit SMILES scratchpad. 
Taken together, the models appear to use their traces as scratchpads in different forms: Chem-R and ether-0 are more strongly SMILES-oriented, whereas ChemDFM-R adopts a more linguistically distributed representation. 
The evidence remains asymmetric. 
Saliency and attention clearly localize the SMILES-oriented components, whereas the positive account of ChemDFM-R's alternative cues rests on broader attribution and entropy patterns because the current token categories do not isolate those cues cleanly.

To summarize, the results support a view of chemical chain-of-thought as a hallucination-prone molecular scratchpad. Its functional-group explanations are often unsupported and weakly coupled to the answer. Fragmented SMILES drafts can influence generation in the Chem-R lineage, attribution suggests that ether-0 also prioritizes SMILES fragments, and ChemDFM-R appears to rely on different candidate cues.
As a proof of concept, continuing Chem-R training with the same verifier as an online process reward reduces mean ER by $73$--$95\%$ across the four task families without a measurable loss on the reported task metrics (cap2mol ER $6.7\to1.8$, retrosynthesis $15.1\to0.8$). 
Chem-R-Faithful also shows greater SMILES-saliency enrichment ($2.5\times\to3.2\times$) and a higher draft-copy rate ($0.14\to0.22$; Fig.~\ref{fig:draft}d), changes that are consistent with the scratchpad account.

%% file: nmi/sections/discussion.tex
\section{Discussion}
\label{sec:discussion}

The experiments reveal a mismatch between what a chemical trace says and which parts of that trace affect the answer. 
Functional-group-level explanations are often unsupported and weakly coupled to generation. 
Structural fragments are less readable as an explanation, but targeted perturbations show that SMILES drafts can affect the output in the Chem-R lineage. 
Chemical CoT is therefore neither a faithful deduction nor entirely post-hoc rationalization. 
It behaves as an externalized workspace for intermediate computation, although the prose in that workspace should not be read as a reliable derivation.

The contrast between Chem-R and ChemDFM-R also matters methodologically. 
Chem-R and ether-0 externalize fragmented SMILES, whereas attribution analyses suggest that ChemDFM-R places more weight on scaffold, positional and nomenclature cues. 
A probe restricted to SMILES would therefore miss candidate signals in ChemDFM-R. 
This model dependence echoes the low consistency reported across chemical representations~\citep{yan2025inconsistency} and argues against assuming a universal CoT format. 
In our analysis, saliency nominates answer-sensitive trace elements and perturbation tests whether changing them affects generation. 
The causal claim for the Chem-R SMILES draft comes from the intervention; attention and saliency provide supporting localization and a descriptive account of the alternative ChemDFM-R cues~\citep{ancona2018towards,jain2019attention}.

This account does not fundamentally conflict with the empirical success of reasoning models in molecular modelling~\citep{wang2025chemr,narayanan2025ether0,li2025molr1,zhao2025molreasoner}.
A trace may improve performance by providing space to draft partial structure even when some of its verbal claims are false. 
Better accuracy, by itself, says little about whether the accompanying rationale is faithful.

In molecular design, retrosynthesis and drug discovery, a natural-language rationale may be the main surface available for human review~\citep{m2024augmenting,mirza2025framework}.
A chemist could reasonably interpret a fabricated group as evidence for the answer, even if the model did not use that claim. We did not test downstream decisions, so this remains a plausible risk rather than a measured harm. For oversight, chemical CoT is better treated as a report whose checkable claims require verification.

The Chem-R-Faithful training experiment also provides a proof of concept for improving faithfulness in molecular reasoning. An answer-only objective cannot distinguish two traces that reach the same answer, even if one contains fabricated claims, consistent with evidence that outcome-based reinforcement learning does not necessarily ensure faithful reasoning traces~\citep{chen2025reasoningmodels}. 
Fabrication is already prevalent before chemistry SFT (Fig.~\ref{fig:pervasive}e). Gating the answer reward on $\mathrm{ER}=0$ provides a way to distinguish otherwise answer-equivalent traces: a correct answer receives no accuracy reward when its reasoning trace fabricates external facts. This connects process supervision~\citep{cobbe2021training,uesato2022solving,lightman2023verify} with domain-grounded verification and tool use in scientific agents~\citep{m2024augmenting,wang2025geneagent}. 
Under this intervention, Chem-R-Faithful produces fewer fabricated claims and places greater emphasis on structural drafts. However, the intervention is not a complete fix: Chem-R-Faithful also writes the complete answer within the trace more often than the released model (draft-copy $0.14\to0.22$; Fig.~\ref{fig:draft}d). 
This increase in early or complete answering may itself constitute a possible shortcut induced by the reward: the model commits to a structure early so that subsequent functional-group claims can be verified against it, rather than necessarily performing more faithful step-by-step reasoning. The reduction in fabrication should therefore be interpreted alongside this behavioural shift, and structure-grounded process rewards warrant explicit monitoring for such shortcuts. Whether the same pattern generalizes to other model families remains open.

\section{Conclusion}

Chemistry provides a uniquely verifiable setting for studying chain-of-thought reasoning because many claims can be checked directly against molecular structures. 
Our results show that unsupported chemical claims are widespread and largely decoupled from answer correctness. 
Mechanistic analyses suggest that readable functional-group claims are often inert, whereas structural drafts can be computationally consequential, supporting the view of chemical CoT as a hallucination-prone molecular scratchpad. 
Chem-R-Faithful further shows that verification-grounded process supervision can reduce such fabrications while preserving task performance. 
Together, these results show that answer correctness and reasoning groundedness are distinct properties of chemical CoT, and that reliable chemical reasoning requires both answer-level evaluation and structure-grounded process verification.

%% file: nmi/sections/methods.tex
\section{Methods}
\label{sec:methods}

\paragraph{Models and tasks.}
We evaluated four off-the-shelf reasoning models: Chem-R-8B~\citep{wang2025chemr}, ChemDFM-R-14B~\citep{zhao2025chemdfm}, ether-0-24B~\citep{narayanan2025ether0} and DeepSeek-R1-Distill-Llama-8B~\citep{guo2025deepseek}. The main evaluation covers twelve generative task variants grouped into four families: caption-to-molecule generation (cap2mol), molecule-to-caption generation (mol2cap), retrosynthesis, and nine S$^2$-Bench structured molecule generation, editing and optimization subtasks. Cap2mol and mol2cap use ChEBI-20~\citep{edwards2022translation}; retrosynthesis uses USPTO-50k~\citep{liu2017retrosynthetic}; S$^2$-Bench is from ~\cite{li2024tomg}. Cap2mol ($n=3,300$), mol2cap ($n=3,300$) and retrosynthesis ($n=5,007$) use the full test sets. Meanwhile, S$^2$-Bench has 9 subtasks and uses first $500$ prompts per subtask for tractability ($n=4,500$). Notably, the evaluation is limited to three established molecular benchmarks, whose task composition and molecular coverage may not represent the full distribution of real-world chemical reasoning problems.

\paragraph{Response parsing.}
Most models delimit reasoning and answer spans with \texttt{<think>}$\ldots$\texttt{</think>} and \texttt{<answer>}$\ldots$\texttt{</answer>} tokens. We normalize equivalent markup before diagnosis, including the delimiter variants used by ether-0, so that all models are scored on a comparable trace and answer span.

\paragraph{Claim extraction and verification.} 
The verifier uses a deterministic rule pipeline. A curated library first maps explicit chemical mentions and their synonyms to functional-group, ring-system or molecular-class categories and identifies transformation statements. Rule-based derivation context then retains claims about the task-relevant molecule while excluding, for example, precursor groups mentioned only as sources of a transformation; this exclusion is applied on caption-to-molecule, where such derivation statements are common, and is off for the other tasks.
Each retained category is assigned an RDKit SMARTS query or an exact structural check~\citep{landrum2013rdkit}. Generic features such as ``ring'' or ``aromatic'' use broad graph predicates, whereas specific groups such as carboxylic acid, ester, phosphate, pyridine or thiophene use tighter substructure definitions. 
The structural verdict uses exact graph matching rather than a generative judge or semantic similarity. The library provides explicit extraction and verification rules for every category evaluated in this study. Phrasings outside this operational vocabulary are not scored, so the reported ER is conditional on the explicit claims recovered by the rule library.
This detector does not certify a complete chemical argument and is less informative for broad mechanistic reasoning, implicit domain knowledge or answers with little molecular structure. We also manually observed $96.9\%$ pooled extraction precision among recovered claims, but the claim-stratified audit does not estimate recall because it did not exhaustively annotate claims missed by the extractor. We accordingly interpret ER as a fabrication rate over the explicit, structurally decidable claims covered by the verifier, rather than as a recall-complete audit of all chemical reasoning in the trace. The audit covered 257 extracted claims from the reported models; a single author with chemistry expertise, blinded to model identity and to the automated verdict, judged each extraction, confirming 249 (96.9\%). The audit does not estimate recall.

\paragraph{Task-aware grounding.}
The grounding set depends on the task. For cap2mol, claims may be grounded by groups named in the input caption, the predicted molecule or the reference molecule.
For mol2cap, claims are checked against the input molecule. For retrosynthesis, claims may refer to the product or predicted reactants. For S$^2$-Bench, the source molecule, instruction-named groups and predicted molecule are all grounding evidence. A claim counts as extrinsic reasoning fabrication only when it is absent from all of the grounding evidence enumerated above for that task.

\paragraph{Hallucination score and derived metrics.}
Each response receives four scores in $[0,100]$ on a $2\times2$ taxonomy that crosses a grounding axis (intrinsic versus extrinsic) with a location axis (reasoning versus output): IR for self-contradiction in the trace, IO for answer invalidity or prompt-constraint failure, ER for fabricated reasoning claims, and EO for output-level structural deviation from the reference or task constraints. The aggregate hallucination score is $0.15\,\mathrm{IR}+0.25\,\mathrm{IO}+0.25\,\mathrm{ER}+0.35\,\mathrm{EO}$, and EO is set to zero on exact-match responses to avoid false positives. When the predicted SMILES cannot be parsed, IO and EO are treated as undefined rather than maximal and the remaining weights are renormalized to sum to one, so such a response is scored $0.375\,\mathrm{IR}+0.625\,\mathrm{ER}$. Claim precision is the fraction of extracted structural claims that verify, and grounded claims is the number of distinct verified specific claims per response; per-claim fabrication rate is the fraction of category-deduplicated claimed groups that are fabricated.

\paragraph{Chemical evidence path.}
Let $x$ be the task input, $T$ the generated reasoning trace, $A$ the generated answer and $\mathcal{K}$ the task-specific chemical knowledge used by the verifier. 
The claim extractor maps the trace to a set of chemical claims $C(T)=\{c_i\}_{i=1}^{m}$. A support function $g(c_i;x,A,\mathcal{K})\in\{0,1\}$ indicates whether claim $c_i$ is grounded by the input, the answer, the reference when available, and the task context. The extrinsic reasoning fabrication score is a bounded penalty that increases with the unsupported-claim rate:
\[
u(T;x,A,\mathcal{K}) = 1-\frac{1}{m}\sum_{i=1}^{m} g(c_i;x,A,\mathcal{K}),
\qquad
\mathrm{ER} = \min\!\left(100,\; \kappa\, u\right)\in[0,100],
\]
where $\kappa$ is a fixed per-task scale ($75$ for cap2mol and $60$ for the remaining tasks), so $\mathrm{ER}$ saturates at $100$ and is not a raw percentage; fabricated molecular-class claims incur an analogous penalty, and responses with no extractable structural claims receive task-specific handling. A trace is clean when $\mathrm{ER}=0$ (equivalently $u=0$). Trace--answer coupling is reported by the clean-minus-fabricating answer-score gap, $\Delta_{\mathrm{couple}}=\mathbb{E}[s(A)\mid \mathrm{ER}=0]-\mathbb{E}[s(A)\mid \mathrm{ER}>0]$, and by the correlation between ER and the task answer score $s(A)$.

\paragraph{Answer-level unidentifiability.}
An answer-only objective has expected reward $J_{\mathrm{ans}}(\pi)=\mathbb{E}_{x,T,A\sim\pi}[r_{\mathrm{ans}}(A,y)]$. Because this reward does not depend on $T$, any two policies with the same answer marginal $\pi(A\mid x)$ obtain the same $J_{\mathrm{ans}}$ even if their expected trace fabrication differs. Answer-level reward therefore cannot distinguish faithful from fabricating traces among answer-equivalent responses; it does not penalize fabricated traces when the final answer remains correct whenever the answer can be produced through an input-to-answer route. A process-level term $J_{\mathrm{proc}}(\pi)=\mathbb{E}[r_{\mathrm{ans}}(A,y)+\lambda\, r_{\mathrm{trace}}(T;x,A,\mathcal{K})]$ separates these otherwise-equivalent traces.

\paragraph{Semantic entropy.}
For semantic entropy~\citep{kuhn2023semantic,farquhar2024detecting} we sample ten responses per prompt at temperature $0.8$, top-$p$ $0.95$, and cluster them by task-aware semantic equivalence: Morgan-fingerprint Tanimoto similarity (threshold $0.85$) for molecule generation, reactant-set equivalence for retrosynthesis, and, for molecule-to-caption, token-overlap agreement between the sampled reasoning traces (threshold $0.80$). Entropy is the Shannon entropy $H=-\sum_c p_c\log p_c$ of the cluster distribution, with $p_c$ the fraction of the ten samples in cluster $c$; it measures confidence, not truth.

\paragraph{Causal perturbation and behavioural drift.}
Two interventions test whether the trace's functional-group claims cause the answer. The first is teacher-forced: on clean cap2mol traces in which a functional-group claim was verified, we replaced one trace mention with an incorrect group and measured $\Delta\log p$, the change in the answer's mean per-token log-probability $\log P(A\mid x,T)/|A|$ under teacher forcing. A synonym replacement is the negative control and the same corruption applied to the input caption is the positive control. The second is behavioural: from a perturbed prefix we regenerate the answer greedily and record the flip-to-wrong rate among originally-correct responses. The conditions are a synonym control, corrupting every functional-group-name claim in the trace, replacing the whole trace with another molecule's real trace, dropping the trace, and corrupting the group in the input. We report $95\%$ normal-approximation (Wald) intervals on the flip rate and paired McNemar tests for the input-versus-trace contrast, and we separate trace claims that are restated from the input from those derived by the model.

\paragraph{Conditional entropy.}
As a metric-free complement, for each example we drew eight answer samples (temperature $0.8$, top-$p$ $0.95$) under three trace conditions, empty, real and functional-group-corrupted, and, for the main models, a fourth condition that swaps in another molecule's whole trace.
Answer entropy is the Shannon entropy over canonical-SMILES clusters. We report information gains relative to the real-trace condition: presence ($H_{\text{noCoT}}-H_{\text{realCoT}}$), content ($H_{\text{corrupt}}-H_{\text{realCoT}}$) and swap ($H_{\text{swap}}-H_{\text{realCoT}}$). This isolates whether corrupting the verbal claims, versus replacing the whole trace, carries answer information without reference to any accuracy metric.

\paragraph{Token-type saliency.}
To localize which trace tokens the answer depends on we use gradient$\times$input saliency~\citep{ancona2018towards} with the model parameters frozen. Let $e_j$ be the input embedding of trace token $j$ and $\log P(A\mid x,T)=\sum_t \log p(a_t\mid a_{<t},x,T)$ the teacher-forced log-probability of the answer $A$ generated by the model, regardless of whether it is correct. We attribute to token $j$ the saliency
\[
  s_j = \bigl|\, e_j \cdot \nabla_{e_j}\log P(A\mid x,T)\,\bigr|,
\]
normalize $s$ to sum to one over the trace tokens, and label each token as a SMILES fragment, functional-group word, position digit, other word or punctuation. For a token type $\tau$ we report its saliency share $\sum_{j\in\tau} s_j$ and its enrichment, the share divided by the fraction of trace tokens of type $\tau$, so a value above one marks above-average answer-sensitivity per token. SMILES fragments are tokens covered by an RDKit-parseable substring of at least six characters; functional-group words are tokens covered by any functional-group synonym in the verifier's library. This probe is run on the full evaluation volume for the Chem-R families and ChemDFM-R ($n=16{,}107$ per model, stratified by ER) and on an $n=1500$ subsample for ether-0.

\paragraph{Attention attribution.}
We treat attention analyses as descriptive rather than causal~\citep{jain2019attention}, use teacher forcing on the full sequence $[\,x \,\|\, \texttt{<think>}\,T\,\texttt{</think>} \,\|\, \texttt{<answer>}\,A\,\texttt{</answer>}\,]$ in a single forward pass (eager attention, bfloat16, no sampling) and read the post-softmax self-attention. For layer $l$ of $L$ with $H$ heads let $A^{(l)}\in[0,1]^{H\times S\times S}$ be the attention and $\bar a^{(l)}=\tfrac1H\sum_{h=1}^{H} A^{(l)}_h$ its head-mean. With $\mathcal A$ the answer-token positions (the span inside \texttt{<answer>}), the answer's attention to a key position $j$, and to a key-token set $\mathcal T$, are the query- and set-averages
\[
  \mathrm{att}^{(l)}(j)=\frac{1}{|\mathcal A|}\sum_{q\in\mathcal A}\bar a^{(l)}_{q,j},
  \qquad
  \mathrm{att}^{(l)}(\mathcal T)=\frac{1}{|\mathcal T|}\sum_{j\in\mathcal T}\mathrm{att}^{(l)}(j).
\]
The per-token (rather than summed) form removes the confound that the trace is several times longer than the input. Every attention statistic is read at the fixed middle layer $l^\star=\lfloor L/2\rfloor$, rather than at a layer selected to maximize the reported contrast. In a layerwise sensitivity check, the input-versus-trace asymmetry has the same direction at every layer except the first and peaks in the middle of the network. Regions are the input-caption span and the \texttt{<think>} trace span, and within each we evaluate $\mathrm{att}(\mathcal T)$ for $\mathcal T$ equal to all tokens, SMILES fragments and functional-group words (defined as above); this region and within-trace salience uses the full evaluation volume ($n=16{,}107$ per model across all Chem-R families, stratified by ER). The matched-token control and the causal perturbation below are reported on cap2mol, the natural-language-input task in which a functional-group word can occur in both the input and the trace: we pick such a word and compare $\mathrm{att}$ at its input occurrence against its trace occurrence; the two occurrences have the same string form, so their ratio isolates location from content and length. The draft-copy rate is the fraction of responses whose RDKit-canonical answer SMILES equals the canonical form of some SMILES substring in the
trace.

\paragraph{Draft-SMILES perturbation.}
To test the drafted structure causally we perturbed only the SMILES a model writes in its trace, leaving functional-group prose intact, on originally-correct examples that actually draft a SMILES. The mask condition replaces the draft with a placeholder (a presence test) and the corrupt condition replaces it with a valid but structurally wrong SMILES (a content/direction test); we compare against corrupting a functional-group name and swapping the whole trace on the same subset. ChemDFM-R, which drafts a SMILES in only $8\%$ of its correct traces, is the negative control.

\paragraph{Process-supervised training.}
Starting from the released Chem-R-8B checkpoint, we continued training with GRPO~\citep{shao2024deepseekmath}. We trained Chem-R-Faithful on $80,000$ examples drawn exclusively from the designated training splits, balanced equally across the four task families with $20,000$ examples per family. No training example overlapped with the evaluation sets. The resulting checkpoint is Chem-R-Faithful (Fig.~\ref{fig:pervasive}e). The reward combines format compliance, the task's answer metric, the anti-hallucination score and a grounded-claim term, with weights $(\lambda_f,\lambda_a,\lambda_h,\lambda_g)=(0.1,0.4,0.4,0.2)$:
\[
\tilde r = \lambda_f r_{\mathrm{format}}+\lambda_a r_{\mathrm{answer}}
  + \lambda_h (1-\mathrm{hallucination}/100) + \lambda_g r_{\mathrm{grounded}}.
\]
We gate the answer-accuracy term on trace faithfulness, paying it only when the trace is clean:
\[
r=\lambda_f r_{\mathrm{format}}+\mathbf{1}[\mathrm{ER}=0]\,\lambda_a r_{\mathrm{answer}}
  +\lambda_h\,(1-\mathrm{hallucination}/100)+\lambda_g r_{\mathrm{grounded}}.
\]
Thus a response with a fabricated reasoning claim earns no accuracy reward, even if its answer is correct, while the format, anti-hallucination and grounded terms are still paid. This prevents the reward hacking in which the policy trades fabrication against answer accuracy or additional claim volume.
The grounded term rewards naming more verified groups while staying precise, $r_{\mathrm{grounded}}=2\,\dfrac{\min(n_{\mathrm{ver}},C)}{C}\,\dfrac{n_{\mathrm{ver}}}{n_{\mathrm{ver}}+n_{\mathrm{fab}}}\in[0,2]$, where $n_{\mathrm{ver}}$ and $n_{\mathrm{fab}}$ count verified and fabricated specific functional groups and the count is capped at $C=5$; groups are de-duplicated by category, so the reward cannot be farmed by repetition. Cap2mol and retrosynthesis use exact match, mol2cap uses caption similarity, and OpenMolIns uses official task success including property-direction success for optimization tasks. We trained one joint model over all four families (twelve variants, 80k balanced examples) and call it Chem-R-Faithful. The same verifier used for evaluation supplies ER, the anti-hallucination score and the grounded-claim counts during training. Chem-R-Faithful is a single training run (936 GRPO steps, data shuffling seed 42) on one model lineage and was not replicated across seeds or model families; we therefore interpret it as a proof of concept.

\paragraph{Training-stage ladder.}
To locate the origin of fabrication, we evaluated matched-prompt generations at four points on a single lineage, each scored on the full task suite (twelve task variants, $n=16{,}107$ responses per stage): the pre-chemistry-SFT base model (Llama-3.1-8B-Instruct), the chemistry SFT model with CoT, the released answer-only GRPO model (Chem-R), and Chem-R-Faithful. For each stage, we measured the per-claim fabrication rate, ER, clean-trace fraction, task performance, and trace-level hedging and abstention rates identified by regular expressions. Figure~\ref{fig:pervasive}e reports the per-claim fabrication rate and task performance along this lineage.

\paragraph{Computational Resources.} All experiments were conducted on NVIDIA H200 GPUs: four GPUs for reinforcement learning and one or two GPUs for inference and the mechanism probes. Training Chem-R-Faithful took 106 GPU-hours. The analysis was inference-dominated rather than training-dominated: it involved 112,749 diagnosed responses (seven models × 16,107 prompts), 384,792 regenerations for the trace perturbation tests, and 593,032 samples for the conditional-entropy and semantic-entropy measures, approximately 1.1 million generations in total. A full twelve-task evaluation sweep for one 8B model took 17–20 GPU-hours on a single GPU.

%% file: nmi/sections/availability.tex
\section*{Data availability}

The benchmark corpora analysed in this study are publicly available from their original sources (ChEBI-20, USPTO-50k and S²-Bench; access routes are documented in data/CORPORA.md of the repository). No new primary dataset was collected. All study-generated data, including evaluation prompts, model responses, parsed reasoning and answer spans, per-response claim-extraction and grounding records, perturbation, conditional-entropy, saliency and attention outputs, the training and validation parquet files, and the source data underlying every figure (data/source\_data.xlsx) are publicly available at \url{https://github.com/phenixace/MolReHallu} under CC BY 4.0; the version corresponding to this paper is release v1.0.0 (commit 0c95196). Redistribution of third-party data and model weights remains subject to their original licences. Source data are provided with this paper.

\section*{Code availability}

The code for reproduction is accessible through \url{https://github.com/phenixace/MolReHallu}. The repository also includes the model and dataset configuration files, dependency specifications, inference and training parameters, and scripts required to reproduce the reported analyses and figures. The version corresponding to this paper is release v1.0.0 (commit 0c95196); the repository includes random seeds, dependency specifications and a verification script that re-derives all diagnosis records and every figure from the released data on a CPU. Chem-R-Faithful is available through \url{https://huggingface.co/phenixace/Chem-R-Faithful}.